\documentclass[prl, twocolumn,superscriptaddress]{revtex4}
\usepackage{graphicx}
\usepackage{amsmath}
\usepackage{amssymb}
\usepackage{bm}
\usepackage{color}
\usepackage{hyperref}
\usepackage{tabularx}
\usepackage{multirow}
\usepackage{braket}

\begin{document}
\title{From Nodal Ring Topological Superfluids to Spiral Majorana Modes in Cold Atomic Systems}
\author{Wen-Yu He}
\affiliation{Department of Physics, Hong Kong University of Science and Technology, Clear Water Bay, Hong Kong, China}
\author{Dong-Hui Xu}
\affiliation{Department of Physics, Hubei University, Wuhan 430062, China}
\author{Benjamin T. Zhou}
\affiliation{Department of Physics, Hong Kong University of Science and Technology, Clear Water Bay, Hong Kong, China}
\author{Qi Zhou}
\affiliation{Department of Physics and Astronomy, Purdue University, 525 Northwestern Avenue, West Lafayette, IN 47907, USA}
\author{K. T. Law} \thanks{phlaw@ust.hk}
\affiliation{Department of Physics, Hong Kong University of Science and Technology, Clear Water Bay, Hong Kong, China}

\begin{abstract}
In this work, we consider a 3D cubic optical lattice composed of coupled 1D wires with 1D spin-orbit coupling. When the s-wave pairing is induced through Feshbach resonance, the system becomes a topological superfluid with ring nodes, which are the ring nodal degeneracies in the bulk, and supports a large number of surface Majorana zero energy modes. The large number of surface Majorana modes remain at zero energy even in the presence of disorder due to the protection from a chiral symmetry. When the chiral symmetry is broken, the system becomes a Weyl topological superfluid with Majorana arcs. With 3D spin-orbit coupling, the Weyl superfluid becomes a novel gapless phase with spiral Majorana modes on the surface. The spatial resolved radio frequency spectroscopy is suggested to detect this novel nodal ring topological superfluid phase.
\end{abstract}

\maketitle

{\emph{Introduction}}--
The study of topological phases has been one of the most important topics in physics in the past decade~\cite{Kane1, Zhang1}. After extensive study of the topological phases which are fully gapped in the bulk, the study of gapless topological phases which possess point nodes has attracted more and more attention~\cite{Wan, Balent1,Luling1, Rappe1, Zhijun1, Zhijun2}. Particularly, 3D Weyl semimetals with topologically protected point nodes and 3D Dirac semimetals with symmetry protected point nodes have been discovered experimentally in condensed matter systems and photonic system~\cite{Liu1, Liu2, Hassan1, Ding, Luling2}. The properties of these materials such as the magneto-resistance are under intense theoretical and experimental studies~\cite{Burkov, Spivak, Genfu, Ong, ZhangC}. Due to the great tunability of optical lattice structures and synthetic gauge fields, fully gapped 1D and 2D topological phases have been realized in shaken optical lattices and Raman optical lattices~\cite{Jotzu, Aidelsburger}. Recently, several proposals have been made to realize nodal Weyl semi-metals~\cite{Jiang, Buljan, Wenyu, Yong1, Burrello, Qi} and Weyl superfluids~\cite{Chuanwei1, Vincent, Chuanwei2}.

More recently, topological phases which possess nodal rings in the bulk have been proposed and studied~\cite{Balent2, Weng, Glatzhofer, Rappe2, Huxiao, ChenFang, Soluyanov, Bian}. Unlike Weyl semimetal phases with surface Fermi arc states which connect the Weyl nodal points in the surface Brillouin zone, topological nodal ring phase supports drumheadlike surface flat bands. Due to the large density of states of the surface states at the Fermi energy, the system is expected to serve as a good platform to study novel phases caused by strong particle-particle interactions. However, it requires rather non-trivial spin-orbit coupling to achieve this nodal ring topological phase.

Unfortunately, due to the complexity of the condensed matter systems, the nodal rings proposed are far from the Fermi energy and they are usually accompanied by bulk states from other bands~\cite{Balent2, Weng, Glatzhofer, Rappe2, Huxiao, ChenFang, Soluyanov, Bian}, which makes the observable feature from the nodal rings ambiguous. On the other hand, cold atoms in optical lattices are clean and have high controllability on both the band structures and interaction strengths. Along with its recent progress in realizing the synthetic spin-orbit coupling~\cite{Spielman, Jingzhang1, Zwierlein, Jingzhang2, Jingzhang3, Jianwei}, cold atom systems are good platforms to simulate and study nodal topological phases.

\begin{figure}
\centering
\includegraphics[width=3.5in]{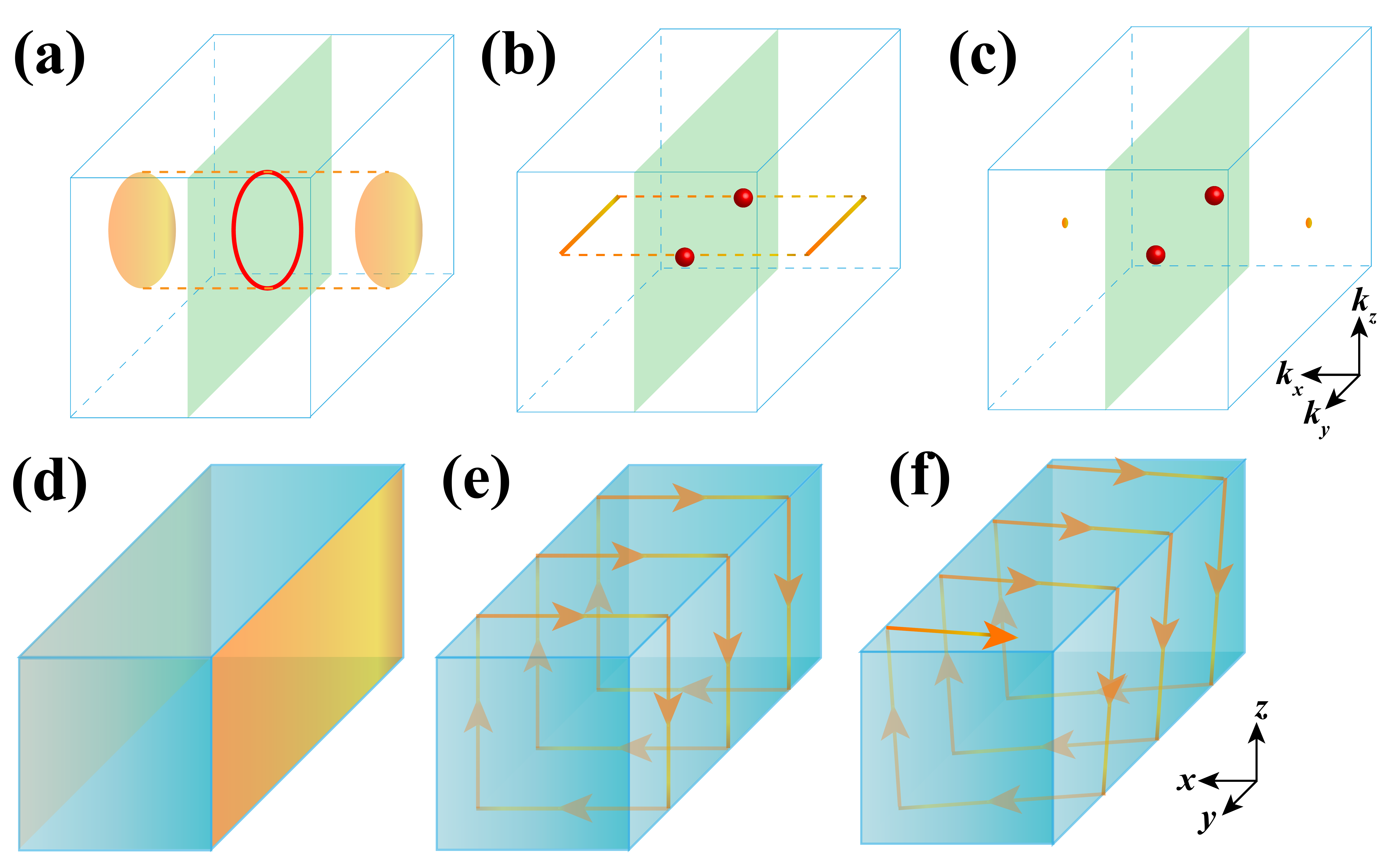}
\caption{(Color online) Schematic illustration of the gapless topological superfluids evolution: (a) ring nodes with 1D spin-orbit coupling and (d) the localized Majorana zero modes, (b) Weyl nodes in zero energy with 2D spin-orbit coupling and (e) the chiral Majorana zero modes, (c) Weyl nodes shifted to different energies with Weyl spin-orbit coupling and (f) the spiral Majorana zero modes. The zero energy spectral function for the surface states is also schematically shown in the surface Brillouin zone. }
\label{Fig1}
\end{figure}

In this work, we study a 3D optical cubic lattice composed of coupled 1D wires with 1D spin-orbit coupling, which have been experimentally realized~\cite{Spielman, Jingzhang1, Zwierlein, Jingzhang2}. Each individual 1D wire has spin-orbit coupling of the form $k_x \sigma_z$ where $k_x$ is the momentum along the wire direction and $\sigma_z$ represents the particle spin.  The wires are coupled through simple hoppings which conserve the particle spin. However, in the presence of on-site attractive interaction, which can be controlled through Feshbach resonance in cold atoms~\cite{Bloch4, Chin}, the system enters the topological superfluid regime with nodal rings. Due to the nodal rings, there exist Majorana pockets with a large number of zero energy Majorana modes in the surface Brillouin zone. The nodal rings realized here are pinned exactly at the Fermi energy and well isolated from other bands. Moreover, the large number of Majorana modes remain at zero energy due to the protection from a chiral symmetry. Interestingly, by breaking the chiral symmetry, most part of the nodal ring is gapped out and the system becomes a Weyl topological superfluid with nodal points and chiral Majorana modes. Further reducing the lattice symmetry by introducing a 3D synthetic spin-orbit coupling of the form $k_x\sigma_z+k_y\sigma_x+k_z\sigma_y$, the Weyl nodes in the Weyl superfluid are shifted in energy and novel spiral Majorana surface states appear. These topological phases are summarised in Fig.\ref{Fig1}. We propose that spatial resolved radio frequency spectroscopy can be used to detect this parent nodal ring topological superfluid phase.

\begin{figure}
\centering
\includegraphics[width=3.5in]{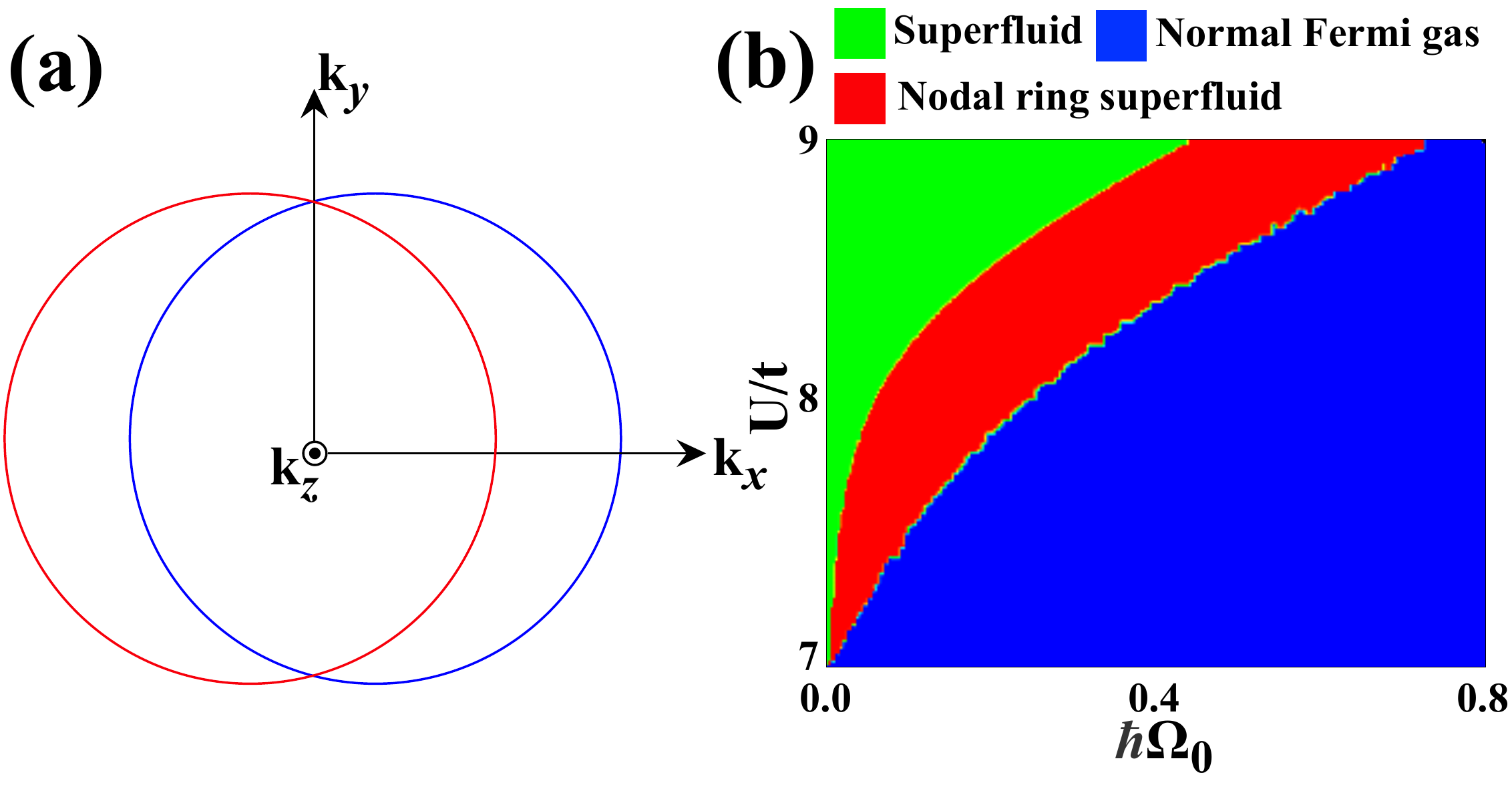}
\caption{(Color online) (a) The Fermi pockets of the spin-orbit coupled Fermi gas with $\mu=0$ in the $\left(k_x, k_y\right)$ plane. The whole Fermi spheres can be obtained through rotating the Fermi pockets around the $k_x$ axis. It is noted that in the $k_x=0$ plane, all states are spin doubly degenerated. (b) The phase diagram for the spin-orbit coupled Fermi gas with $\mu=0$ and $\lambda_{\textrm{SOC}}=0.25t$. The nodal ring superfluid phase is sandwiched between the normal superfluid phase and the normal Fermi gas.}
\label{Fig2}
\end{figure}

{\emph{The scheme}}--
We consider the two magnetic sublevels $\left |\uparrow \right \rangle=\left |F=9/2, m_F=-7/2\right \rangle$ and $\left |\downarrow \right \rangle=\left |F=9/2, m_F=-9/2\right \rangle$ in $^{40}\textrm{K}$ atoms as the two spin-$1/2$ states. A pair of Raman lasers counter propagating along $x$ direction with frequencies $\omega_1$, $\omega_2$ and wave vectors $\bm{k}_1$, $\bm{k}_2$ are applied to couple the two spin states through the two-photon Raman transition and generate the effective 1D spin-orbit coupled single particle Hamiltonian $\tilde{H}_0=\frac{\left(\bm{p}-\bm{q}\sigma_z\right)^2}{2m}+\frac{\delta}{2}\sigma_z+\hbar\Omega\sigma_x$~\cite{Spielman, Jingzhang1, Zwierlein, Jingzhang2}, where $\bm{p}=-i\hbar\nabla$, $m$ is the atomic mass, $\delta$ is the detuning, $\Omega$ is referred to the two-photon Rabi frequency, and $\bm{q}=\frac{1}{2}\left(\bm{k}_1-\bm{k}_2\right)$ is the momentum transfer between the atoms and the Raman laser. The heating caused by the Raman lasers can be reduced through using a relative fast procedure to load  fermions into the lowest Raman-dressed band~\cite{Spielman2}. In the resonant Raman process, we consider the detuning $\delta\approx0$. A cubic lattice potential $V= E_{\textrm{r}}\left(\nu_x\cos^2Kx+\nu_yV_y\cos^2Ky+\nu_zV_z\cos^2Kz\right)$ is further introduced to trap the $^{40}\textrm{K}$ atoms into the cubic optical lattice, where the $E_{\textrm{r}}=\frac{\hbar^2K^2}{2m}$ is the recoil energy and $\nu_{x, y, z}$ characterizes the depth of the optical lattice potential. For this 1D spin-orbit coupled system in the cubic lattice, a tight binding model is constructed~\cite{Supplementary}. We define the annihilation (creation) operator of $^{40}\textrm{K}$ atoms with spin $s$ as $c^{\left(\dagger\right)}_{s}$ where $s=\uparrow/\downarrow$. Then in the basis of $\left[c_{\bm{k}\uparrow}, c_{\bm{k}\downarrow}\right]^{\textrm{T}}$, the tight binding Hamiltonian in the $\bm{k}$ space can be written as
\begin{align}\label{TB1}
\mathcal{H}_0\left(\bm{k}\right)&=\xi\left(\bm{k}\right)+\hbar\Omega_0\sigma_x+2\lambda_{\textrm{SOC}}\sin k_xa\sigma_z
\end{align}
where $\xi\left(\bm{k}\right)=2t\left(3-\cos k_xa-\cos k_ya-\cos k_za\right)$ and $\sigma$ is the Pauli matrix in the spin space. Here $t$ and $\lambda_{\textrm{SOC}}$ characterize respectively the spin-independent and spin-dependent nearest neighbor hopping in the cubic optical lattice with lattice constant $a=\frac{\pi}{K}$, and $\Omega_0$ is the effective strength of Raman coupling that acts as a Zeeman field in $x$ direction. This 1D spin-orbit coupling can be regarded as an effective Zeeman field in the momentum space, which shifts the Fermi spheres of the spin up and spin down states away from each other along $k_x$ direction and the spin doubly degenerated states only survive in the $k_x=0$ plane, as is seen from Fig. \ref{Fig2}(a). 

We next consider the spin-orbit coupled Fermi gas with s-wave contact interaction, which can be well controlled with the Feshbach resonance. In the optical lattice, this s-wave contact interaction can be described as $H_{\textrm{int}}=-\sum_{i}Un_{i\uparrow}n_{i\downarrow}$, where $U$ is the onsite attractive interaction strength and $n_{is}$ is the density operator for particles with spin $s$ at lattice site $i$. For the magnetic sublevels $\left |\uparrow \right \rangle=\left |F=9/2, m_F=-7/2\right \rangle$ and $\left |\downarrow \right \rangle=\left |F=9/2, m_F=-9/2\right \rangle$ of $^{40}\textrm{K}$ atoms, these hyperfine states exhibit a broad Feshbach resonance at 202.1 G with width $\simeq7G$~\cite{Jingzhang2, Jin, Rosch, Spielman2}. This s-wave Feshbach resonance enables the two-component fermionic atoms to form the singlet BCS cooper pairs and enter the superfluid phase~\cite{Jin}. Under the mean field approximation, the s-wave superfluid order parameter is defined as $\Delta=-U/A\sum_{\bm{k}}\left \langle c_{\bm{k}\downarrow}c_{-\bm{k}\uparrow}\right \rangle$ with $U>0$ and $A$ the volume of optical lattice. The detail about the mean-field solution for $\Delta$ can be found in the Supplementary Information~\cite{Supplementary}. Then the Bogliubov-de Gennes Hamiltonian in Nambu spinor basis $\left[c_{\bm{k},\uparrow}, c_{\bm{k},\downarrow}, c_{\bm{-k},\uparrow}, c_{\bm{-k},\downarrow}\right]$ can be written as
\begin{align}\label{TB2}\nonumber
\mathcal{H}\left(\bm{k}\right)&=\left[\xi\left(\bm{k}\right)-\mu\right]\tau_z+\hbar\Omega_0\tau_z\sigma_x\\
&+2\lambda_{\textrm{SOC}}\sin k_xa\sigma_z+\Delta\tau_y\sigma_y.
\end{align}
Here $\mu$ is the chemical potential and $\tau$ is the Pauli matrix in the Nambu space.

In the 1D spin-orbit coupled Fermi gas, once it enters the superfluid phase, the BCS cooper pairs condensate and the pairing gap is formed at the Fermi level. In the presence of the Zeeman field $\hbar\Omega_0$, the atoms tend to align the spin anti-parallel to the external Zeeman field and this magnetization would compete with the BCS cooper pair condensation. At $k_x=0$ plane, the spin-orbit coupling field is zero and the atoms with $\left | \uparrow \right \rangle$ states and $\left | \downarrow \right \rangle$ states cannot pair once the Zeeman field $\hbar\Omega_0$ exceeds $\Delta$ and as a result the Bogliubov-de Gennes quasiparticle spectrum becomes gapless. Away from the plane $k_x=0$, the spin-orbit coupling acts as a momentum dependent Zeeman field and pins the time reversal partners to $\left | \uparrow \right \rangle$ and $\left | \downarrow \right \rangle$ states respectively, which favors the BCS pairing mechanism. Due to the spin-orbit coupling, the external in-plane Zeeman field $\Omega$ becomes less effective to align the spins anti-parallel along its direction and thus the pairing gap remains finite. Consequently, the in-plane upper critical Zeeman field is strongly enhanced by the spin-orbit coupling~\cite{Ye, Fai}. Importantly, a nodal superfluid phase emerges once the Zeeman field $\hbar\Omega_0$ goes beyond $\Delta$, as is seen from the phase diagram in Fig.\ref{Fig2}(b).

\begin{figure}
\centering
\includegraphics[width=3.5in]{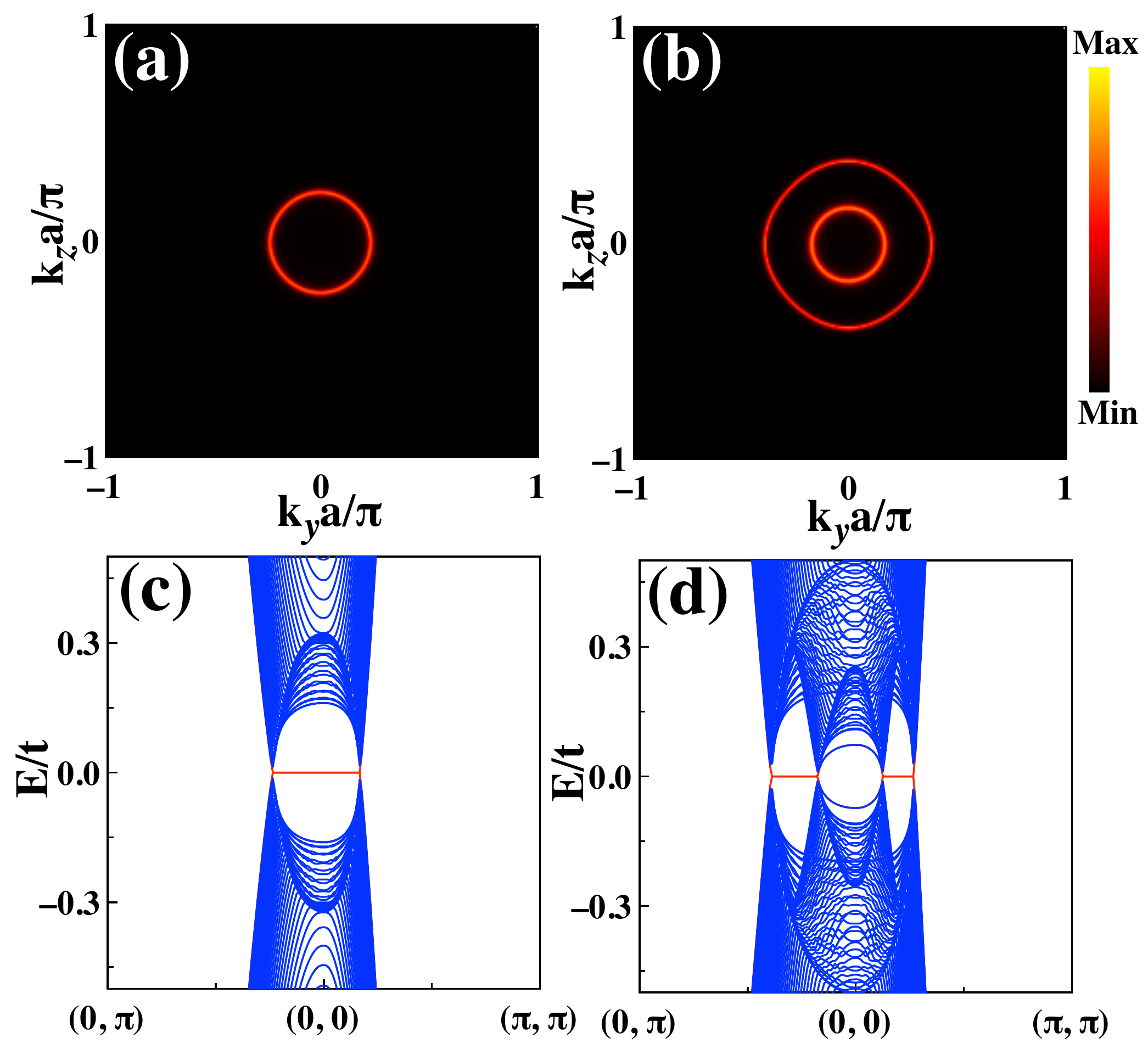}
\caption{(Color online) The ring nodes in the superfluid phase with $U=9t$ and $\hbar\Omega_0=0.6t$. The spectral function at the zero energy for the superfluid shows the single ring node (a) and the double ring nodes (b) in the $k_x=0$ plane. Along the path $\left(k_ya, k_za\right)=\left(0, \pi\right)\rightarrow\left(0, 0\right)\rightarrow\left(\pi, \pi\right)$ of the surface Brillouin zone, the quasi-energy edge spectrum (c) for the single ring node and (d) for the double ring nodes is plotted. The states in the flat band compose of the surface Majorana pocket. }
\label{Fig3}
\end{figure}

\begin{figure}
\centering
\includegraphics[width=3.5in]{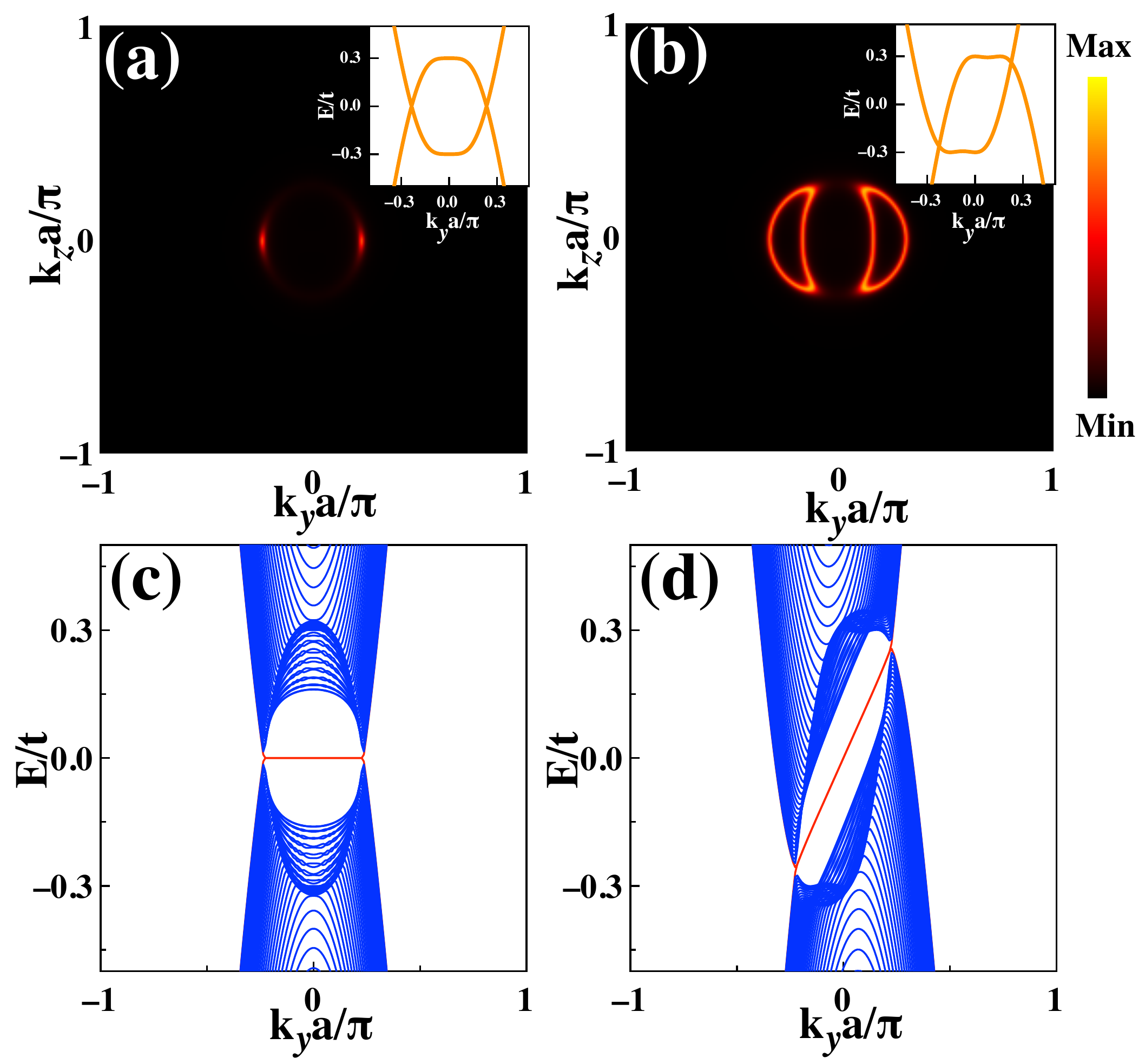}
\caption{(Color online) The Weyl nodes evolved from the single ring node in the superfluid phase with $U=9t$ and $\hbar\Omega_0=0.6t$. The spectral function at the zero energy in the $k_x=0$ plane is plotted for (a) the Weyl superfluid with the Weyl nodes at the same energy and (b) the Weyl superfluid with the energy shifted Weyl nodes. The insets in (a) and (b) correspond to the bulk energy spectrum crossing the Weyl nodes. In the surface Brillouin zone, the drumhead like Majorana edge states are reduced to the Majorana arc. The Majorana arc is the flat band in (c) for the Weyl nodes with equal energy while it supports the unidirectional Majorana edge states in (d) when the Weyl nodes are energy shifted.}
\label{Fig4}
\end{figure}

{\emph{Nodal ring superfluid}}--
The BdG Hamiltonian for the superfluid phase of the 1D spin-orbit coupled Fermi gas can be mapped to the atomic quantum wire that hosts Majorana end states~\cite{Sato, Oppen, Sarma, Zoller}. Since the atomic spins are only coupled to $k_x$, for fixed $k_y$ and $k_z$ the Hamiltonian is effectively reduced to $\mathcal{H}_{k_y, k_z}\left(k_x\right)$. It enters topological nontrivial phase when the criterion $\hbar^2\Omega_0^2>\mu'^2+\Delta^2$ is satisfied~\cite{Sato}, where $\mu'$ is the modified chemical potential $\mu'=\mu-2t\left(2-\cos k_ya-\cos k_za\right)$. To illustrate the emergence of the nodal ring, the spectral function $A\left(E, \bm{k}\right)=-\frac{1}{\pi}\textrm{Tr}\left[\textrm{Im}G\left(E, \bm{k}\right)\right]$, where $G\left(E, \bm{k}\right)=\left(E+i\eta-\mathcal{H}\right)^{-1}$ is the retarded Green's function, is shown in Fig.\ref{Fig3} with $E=0$. In the case $\hbar\Omega_0>\Delta$ and $\mu=0$, the topological nontrivial region emerges in the $\bm{k}$ space with a nodal ring $2t\left(2-\cos k_y-\cos k_z\right)=\sqrt{\hbar^2\Omega_0^2-\Delta^2}$ manifesting the topological transition, as is seen from Fig.\ref{Fig3}(a). Gradually increasing the chemical potential $\mu$ from 0 to $\mu=4t$, another nodal ring emerges once $\mu>\sqrt{\hbar^2\Omega_0^2-\Delta^2}$ and the topological nontrivial region is between the outer ring node $2t\left(2-\cos k_y-\cos k_z\right)=\mu+\sqrt{\hbar^2\Omega_0^2-\Delta^2}$ and the inner ring node $\mu-\sqrt{\hbar^2\Omega_0^2-\Delta^2}=2t\left(2-\cos k_y-\cos k_z\right)$, as seen from Fig.\ref{Fig3}(b). As the Hamiltonian $\mathcal{H}$ respects a time-reversal like symmetry $U_T\mathcal{H}\left(k_x, k_y, k_z\right)U_T^{-1}=\mathcal{H}\left(-k_x, k_y, k_z\right)$ with $U_T=-i\tau_z\sigma_x K$ and the 1D particle-hole symmetry $U_{\textrm{P}}\mathcal{H}\left(k_x, k_y, k_z\right)U_{\textrm{P}}^{-1}=-\mathcal{H}\left(-k_x, k_y, k_z\right)$ with $U_{\textrm{P}}=\tau_xK$. For fixed $k_y$ and $k_z$ $\mathcal{H}$ respects the chiral symmetry $C\mathcal{H}_{k_y, k_z}\left(k_x\right)C^{-1}=-\mathcal{H}_{k_y, k_z}\left(k_x\right)$ with $C=U_{\textrm{P}}U_T=\tau_y\sigma_x$. As a result, the Hamiltonian $\mathcal{H}_{k_y, k_z}\left(k_x\right)$ belongs to the BDI class and the topological invariant is simply the winding number~\cite{Tewari}.

In the $\bm{k}$ space, for the range of $k_y$ and $k_z$ satisfying $\mu-\sqrt{\hbar^2\Omega_0^2-\Delta^2}<2t\left(2-\cos k_y-\cos k_z\right)<\mu+\sqrt{\hbar^2\Omega_0^2-\Delta^2}$ the winding number $N_{\textrm{BDI}}$ is nonzero, as is calculated in the Supplementary Information~\cite{Supplementary}. As a result, in the surface Brillouin zone, the region with nonzero winding number $N_{\textrm{BDI}}$ is filled with a pocket of zero energy states, as seen from Fig.\ref{Fig3} (c) and (d). The zero energy states are well localized on the surface of the system schematically shown in Fig.\ref{Fig1}(d) and satisfy the Majorana criterion $\gamma\left(k_y, k_z\right)=\gamma^{\dagger}\left(-k_y, -k_z\right)$, so they form the Majorana pocket. Due to the chiral symmetry, the Majorana zero modes on each side have definite chirality $\hat{C}\gamma_{\textrm{L}/\textrm{R}}\left(k_y, k_z\right)=\mp\gamma_{\textrm{L}/\textrm{R}}\left(k_y, k_z\right)$, where the chiral operator reads $\hat{C}_{i, i}=C$ at the lattice site $i$ and L/R indicates the Majorana zero modes on the left/right side surface. The Majorana zero modes are well protected by the chiral symmetry as the number of stable Majorana zero modes is equivalent to the net chirality number on the surface~\cite{Yokoyama}. Since the onsite disorder does not break the chiral symmetry, the Majorana zero modes are even robust against disorder~\cite{Yokoyama, Law1}.

{\emph{Evolution to Weyl superfluid}}--
The topological nodal ring superfluid is protected by the chiral symmetry. When another component of spin-orbit coupling term $k_z\sigma_y$ is introduced to form the 2D synthetic spin-orbit coupling, the chiral symmetry is broken. Now at the plane $k_x=0$, for nonzero $k_z$, the new spin-orbit coupling field $k_z\sigma_y$ aligns atomic spins at $\bm{k}$ and $-\bm{k}$ to opposite directions so that they can form Cooper pairs, while at the $k_z=0$ point, the spin-orbit coupling field vanishes and so does the pairing gap. As a result, the new spin-orbit coupling field lifts the ring nodal degeneracies while the point nodes remain degenerate. In this case, the topological nodal ring superfluid is further driven into the Weyl superfluid phase, as is shown in Fig.\ref{Fig4}(a). With the recent experimental progress in synthetic 2D spin-orbit coupling~\cite{Jingzhang3, Jianwei}, this Weyl superfluid phase is promising to be realized. In the presence of the 2D spin-orbit coupling, the Bogliubov-de Gennes Hamiltonian $\mathcal{H}_1$ becomes 
\begin{align}\nonumber
\mathcal{H}_1\left(\bm{k}\right)&=\left[\xi\left(\bm{k}\right)-\mu\right]\tau_z+\hbar\Omega_0\tau_z\sigma_x+\Delta\tau_y\sigma_y\\
&+2\lambda_{\textrm{SOC}}\left(\sin k_xa\sigma_z+\sin k_za\tau_z\sigma_y\right).
\end{align}
Since the atomic spin states $\left | \uparrow \right\rangle$ and $\left | \downarrow \right\rangle$ couple with both $k_x$ and $k_z$, in the $\bm{k}$ space, the 2D spin-orbit coupling produces helical spin texture in the $k_x$ and $k_z$ plane, and the chemical potential is effectively modified by the $k_y$ terms in $\mathcal{H}_1$ as: $\mu'=\mu-2t\left(1-\cos k_ya\right)$. As a result, the whole system sliced by $k_y$ is equivalent to the topological superfluid that supports the chiral Majorana edge states when $\hbar^2\Omega_0^2>\mu'^2+\Delta^2$~\cite{Sato}. Parametrized by $k_y$ the Bogliubov-de Gennes quasiparticle spectrum inevitably undergoes band crossings which lead to topological phase transitions. In particular, when $\mu<\sqrt{\hbar^2\Omega_0^2-\Delta^2}$ there exists one pair of Weyl nodes, while for $\mu>\sqrt{\hbar^2\Omega_0^2-\Delta^2}$ two pairs of Weyl nodes can emerge. The Weyl nodes in the $k_y$ axis make a sharp boundary between the topological trivial and nontrivial segment. In the nontrivial segment, there are chiral Majorana zero modes circulating on the surface seen from Fig.\ref{Fig1}(e). In the surface Brillouin zone, the Weyl nodes are connected by the Majorana arcs, as is seen from Fig.\ref{Fig4}(c).

{\emph{Spiral Majorana zero modes}}--
The Weyl nodes formed in the 2D spin-orbit coupled superfluid phase are very stable and cannot be removed unless mutual annihilation between Weyl nodes of opposite chiralities occur due to strong perturbations. Weak perturbations can only shift the Weyl nodes in the momentum and energy space. When the 2D spin-orbit coupling evolves to the 3D synthetic spin-orbit coupling $\left(k_x\sigma_z+k_y\sigma_x+k_z\sigma_y\right)$, which is also known as the Weyl spin-orbit coupling~\cite{Spielman3, Juzeliunas, Xiongjun}, the Weyl nodes in the Bogliubov-de Gennes quasiparticle spectrum are shifted to different energies as seen from Fig.\ref{Fig4}(b). The corresponding Bogliubov-de Gennes Hamiltonian $\mathcal{H}_2$ reads
\begin{align}\nonumber
\mathcal{H}_2\left(\bm{k}\right)&=\left[\xi\left(\bm{k}\right)-\mu\right]\tau_z+\hbar\Omega_0\tau_z\sigma_x+\Delta\tau_y\sigma_y\\
&+2\lambda_{\textrm{SOC}}\left(\sin k_xa\sigma_z+\sin k_ya\sigma_x+\sin k_za\tau_z\sigma_y\right).
\end{align}
At $k_x=0$, $k_z=0$, the Bogliubov-de Gennes quasi-particle spectrum can be solved to be $E=\pm\sqrt{\Delta^2+\left[2\lambda_{\textrm{SOC}}\sin k_y-\xi\left(\bm{k}\right)+\mu\right]^2}\pm\hbar\Omega_0$. Clearly the spin-orbit coupling field in the $k_y$ direction shifts the Weyl nodes to different energies. Consequently, the Majorana arc connecting the two Weyl nodes of opposite chiralities acquires a finite slope shown in Fig. \ref{Fig4}(d). On the surface, now the circulating Majorana zero modes obtain nonzero group velocity along $y$ direction, so the Majorana zero modes spiral forward on the surface of the system seen in Fig. \ref{Fig1}(f). Meanwhile, a back-flowing current in the bulk will compensate the edge currents induced by these spiral Majorana zero modes~\cite{Law1}.

{\emph{Detection}}--
Characterized by the drumhead like edge states with large density of states, the topological nodal ring superfluid can manifest itself through local density of states measurements. In ultracold Fermi gases, the spatial resolved radio frequency spectroscopy which provides local information has been experimentally implemented~\cite{Shin} and it is considered to detect the large local density of states at the edge~\cite{Pu, Zoller, Mueller}. Experimentally, radio frequency field can be applied to transfer the $\left |\uparrow \right \rangle=\left |F=9/2, m_F=-7/2\right \rangle$ fermions to an unoccupied third hyperfine state $\left |3\right \rangle=\left |F=9/2, m_F=-5/2\right \rangle$, and then the population change in the final state can be measured to represent the spectral response, which can be approximated from the linear response theory
\begin{align}\nonumber
I\left(\bm{r}\right)&=\frac{d}{dt}\left \langle c^{\dagger}_{\bm{r}\uparrow}c_{\bm{r}\uparrow}\right \rangle\\
&\propto \rho_{\uparrow}\left(\bm{r},-\mu-\nu+\varepsilon_3\right)n_{\textrm{F}}\left(-\mu-\nu+\varepsilon_3\right),
\end{align}
where $\rho_{\uparrow}$ is the single particle spectral function, $n_{\textrm{F}}$ is the Fermi distribution, $\varepsilon_3$ denotes the eigen-energy for the third state and $\nu$ the radio frequency detuning. Here the final-state interaction in $^{40}\textrm{K}$ atoms is typically small and can be neglected~\cite{Bloch4}. With the radio frequency detuning $\nu=\varepsilon_3-\mu$, the large number of zero energy edge states in topological nodal ring superfluid phase are expected to greatly enhance the population transfer to the final state and thus can serve as a signature of the topological phase.

{\emph{Conclusion}}--
In conclusion, we propose the realization of topological nodal ring superfluid phase in spin-orbit coupled Fermi gas. The nodal ring superfluid can further evolve into the Weyl superfluid with the synthetic 2D spin-orbit coupling. Involved with the synthetic 3D spin-orbit coupling, the Weyl nodes in the superfluid are shifted in the energy and the system possesses the novel spiral Majorana zero modes on the surface. Our study of the topological nodal superfluid is expected to provide a cold atomic platform to study the creation and manipulation~\cite{Baranov, Buchler} of multiple Majorana states in nodal systems.

{\emph{Acknowledgement}}--
The authors thank the support of HKRGC through HKUST3/CRF/13G and C6026-16W. K. T. L. is further supported by the Croucher Foundation and the Dr Tai-chin Lo Foundation. Q. Z. acknowledges the startup funds from Purdue University. D.-H. Xu is supported by the National Natural Science Foundation of China (under Grant No. 11704106) and the Chutian Scholars Program in Hubei Province. W.-Y. He acknowledges the support of Hong Kong PhD Fellowship.


\begin{thebibliography}{99}

\bibitem{Kane1} M. €‰Z. Hasan and C. €‰L. Kane, Rev. Mod. Phys. {\bf82}, 3045 (2010).

\bibitem{Zhang1} Xiao-Liang Qi and Shou-Cheng Zhang, Rev. Mod. Phys. {\bf83}, 1057 (2011).

\bibitem{Wan} Xiangang Wan, Ari M. Turner, Ashvin Vishwanath, and Sergey Y. Savrasov, Phys. Rev. B {\bf83}, 205101 (2011).

\bibitem{Balent1} A. A. Burkov and Leon Balents, Phys. Rev. Lett. {\bf107}, 127205 (2011).

\bibitem{Luling1} L. Lu, L. Fu, J. D. Joannopoulos, and M. Soljacic, Nat. Photonics {\bf 7}, 294 (2013).

\bibitem{Rappe1} S. M. Young, S. Zaheer, J. C. Y. Teo, C. L. Kane, E. J. Mele, and A. M. Rappe, Phys. Rev. Lett. {\bf108}, 140405 (2012).

\bibitem{Zhijun1} Z. Wang, Y. Sun, X.-Qiu Chen,  C. Franchini, G. Xu, H. Weng, X. Dai, and Z. Fang, Phys. Rev. B {\bf85}, 195320 (2012).

\bibitem{Zhijun2} Z. Wang, H. Weng, Q. Wu, X. Dai, and Z. Fang, Phys. Rev. B {\bf88}, 125427 (2013).

\bibitem{Liu1} Z. €‰K. Liu et al., Science {\bf343}, 864 (2014).

\bibitem{Liu2} Z. €‰K. Liu et al., Nat. Mater. {\bf13}, 677 (2014).

\bibitem{Hassan1} S.-Y. Xu et al., Science {\bf349}, 613 (2015).

\bibitem{Ding} B. Q. Lv, H. M. Weng, B. B. Fu, X. P. Wang, H. Miao, J. Ma, P. Richard, X. C. Huang, L. X. Zhao, G. F. Chen, Z. Fang, X. Dai, T. Qian, and H. Ding, Phys. Rev. X {\bf5}, 031013 (2015).

\bibitem{Luling2} L. Lu, Z. Wang, D. Ye, L. Ran, L. Fu, J. D. Joannopoulos, and M. Soljacic, Science {\bf 349}, 622 (2015).

\bibitem{Burkov} A. A. Zyuzin and A. A. Burkov, Phys. Rev. B {\bf86}, 115133 (2012).

\bibitem{Spivak} D. T. Son and B. Z. Spivak, Phys. Rev. B {\bf88}, 104412 (2013).

\bibitem{Genfu} X. Huang, L. Zhao, Y. Long, P. Wang, D. Chen, Z. Yang, H. Liang, M. Xue, H. Weng, Z. Fang, X. Dai, and G. Chen, Phys. Rev. X {\bf5}, 031023 (2015).

\bibitem{Ong} J. Xiong, S. €‰K. Kushwaha, T. Liang, J. €‰W. Krizan, M. Hirschberger, W. Wang, R. €‰J. Cava, and N. €‰P. Ong, Science {\bf350}, 413 (2015).

\bibitem{ZhangC} C. Zhang, S.-Y. Xu, I. Belopolski, Z. Yuan, Z. Lin, B. Tong, N. Alidoust, C.-C. Lee, S.-M. Huang, H. Lin et al., Nat. Commun. {\bf7}, 10735 (2016).

\bibitem{Jotzu} G. Jotzu, M. Messer, R. Desbuquois, M. Lebrat, T. Uehlinger, D. Greif, and T. Esslinger, Nature (London) {\bf515}, 237 (2014).

\bibitem{Aidelsburger} M. Aidelsburger, M. Lohse, C. Schweizer, M. Atala, J. €‰T. Barreiro, S. Nascimbene, N. €‰R. Cooper, I. Bloch, and N. Goldman, Nat. Phys. {\bf11}, 162 (2015).

\bibitem{Jiang} J. H. Jiang, Phys. Rev. A {\bf85}, 033640 (2012).

\bibitem{Buljan} T. Dubcek, C. J. Kennedy, L. Lu, W. Ketterle, M. Soljacic, and H. Buljan, Phys. Rev. Lett. {\bf114}, 225301 (2015).

\bibitem{Wenyu} W.-Y. He, S. Zhang, and K. T. Law, Phys. Rev. A {\bf94}, 013606 (2016).

\bibitem{Yong1}  Y. Xu and L.-M. Duan, Phys. Rev. A {\bf94}, 053619 (2016).

\bibitem{Burrello} L. Lepori, I. C. Fulga, A. Trombettoni, and M. Burrello, Phys. Rev. A {\bf94} 053633 (2016).

\bibitem{Qi} L.-J. Lang, S.-L. Zhang, K. T. Law and Q. Zhou, Phys. Rev. A {\bf95} 053615 (2017).

\bibitem{Chuanwei1} Y. Xu, R.-L. Chu, and C. Zhang, Phys. Rev. Lett. {\bf112}, 136402 (2014).

\bibitem{Vincent} B. Liu, X. Li, L. Yin, and W. V. Liu, Phys. Rev. Lett. {\bf114}, 045302 (2015).

\bibitem{Chuanwei2} Y. Xu, F. Zhang, and C. Zhang, Phys. Rev. Lett. {\bf115}, 265304 (2015).

\bibitem{Balent2} A. A. Burkov, M. D. Hook, and Leon Balents, Phys. Rev. B {\bf84}, 235126 (2011).

\bibitem{Weng} Hongming Weng, Yunye Liang, Qiunan Xu, Rui Yu, Zhong Fang, Xi Dai, and Yoshiyuki Kawazoe, Phys. Rev. B {\bf92}, 045108 (2015).

\bibitem{Glatzhofer} Kieran Mullen, Bruno Uchoa, and Daniel T. Glatzhofer, Phys. Rev. Lett. {\bf115}, 026403 (2015).

\bibitem{Rappe2} Youngkuk Kim, Benjamin J. Wieder, C. €‰L. Kane, and Andrew M. Rappe, Phys. Rev. Lett. {\bf115}, 036806 (2015).

\bibitem{Huxiao} Rui Yu, Hongming Weng, Zhong Fang, Xi Dai, and Xiao Hu, Phys. Rev. Lett. {\bf115}, 036807 (2015).

\bibitem{ChenFang} C. Fang, Y. Chen, H.-Y. Kee, and L. Fu, Phys. Rev. B {\bf92}, 081201(R) (2015).

\bibitem{Soluyanov} T. Bzdusek, Q. Wu, A. Ruegg, M. Sigrist, and A. A. Soluyanov, Nature {\bf538}, 75 (2016).

\bibitem{Bian} G. Bian, T.-R. Chang, R. Sankar, S.-Y. Xu, H. Zheng, T. Neupert, C.-K. Chiu, S.-M. Huang, G. Chang, and I. Belopolski et al., Nat. Commun. {\bf7}, 10556 (2016).

\bibitem{Spielman} Y.-J. Lin, K. Jimenez-Carcia, and I. €‰B. Spielman, Nature (London) {\bf471}, 83 (2011).

\bibitem{Jingzhang1} P. Wang, Z.-Q. Yu, Z. Fu, J. Miao, L. Huang, S. Chai, H. Zhai, and J. Zhang, Phys. Rev. Lett. {\bf109}, 095301 (2012).

\bibitem{Zwierlein} L. €‰W. Cheuk, A. €‰T. Sommer, Z. Hadzibabic, T. Yefsah, W. €‰S. Bakr, and M. €‰W. Zwierlein, Phys. Rev. Lett. {\bf109}, 095302 (2012).

\bibitem{Jingzhang2} Z. Fu, L. Huang, Z. Meng, P. Wang, L. Zhang, S. Zhang, H. Zhai, P. Zhang, and J. Zhang, Nat. Phys. {\bf10}, 110 (2014).

\bibitem{Bloch4} I. Bloch, J. Dalibard, and W. Zwerger, Rev. Mod. Phys. {\bf80}, 885 (2008).

\bibitem{Chin} C. Chin, R. Grimm, P. Julienne, and E. Tiesinga, Rev. Mod. Phys. {\bf82}, 1225 (2010).

\bibitem{Jingzhang3} L. Huang, Z. Meng, P. Wang, P. Peng, S.-L. Zhang, L. Chen, D. Li, Q. Zhou, and J. Zhang, Nat. Phys. {\bf10}, 588 (2014).

\bibitem{Jianwei} Z. Wu, L. Zhang, W. Sun, X.-T. Xu, B.-Z. Wang, S.-C. Ji, Y. Deng, S. Chen, X.-J. Liu, and J.-W. Pan, Science {\bf54}, 83 (2016).

\bibitem{Spielman2} R. €‰A. Williams, M. €‰C. Beeler, L. €‰J. LeBlanc, K. Jimenez-Garcia, and I. €‰B. Spielman, Phys. Rev. Lett. {\bf111}, 095301 (2013).

\bibitem{Supplementary} See the Supplementary Information.

\bibitem{Jin} C. A. Regal, M. Greiner, and D. €‰S. Jin, Phys. Rev. Lett. {\bf92}, 040403 (2004).

\bibitem{Rosch} U. Schneider, L. Hackermuller, J. €‰P. Ronzheimer, S. Will, S. Braun, T. Best, I. Bloch, E. Demler, S. Mandt, D. Rasch, and A. Rosch, Nat. Phys. {\bf8}, 213 (2012).

\bibitem{Ye} J. M. Lu, O. Zheliuk, I. Leermakers, N. F. Q. Yuan, U. Zeitler, K. T. Law, and J. T. Ye, Science {\bf350}, 1353 (2015).

\bibitem{Fai} X. X. Xi, Z. F. Wang, W. W. Zhao, J. H. Park, K. T. Law, H. Berger, L. Forro, J. Shan, and K. F. Mak, Nat. Phys. {\bf12}, 139 (2016).

\bibitem{Sato} M. Sato, Y. Takahashi, and S. Fujimoto, Phys. Rev. Lett. {\bf103}, 020401 (2009).

\bibitem{Oppen} Y. Oreg, G. Refael, and F. von Oppen, Phys. Rev. Lett. {\bf105}, 177002 (2010).

\bibitem{Sarma} J. €‰D. Sau, R. M. Lutchyn, S. Tewari, and S. Das Sarma, Phys. Rev. Lett. {\bf104}, 040502 (2010).

\bibitem{Zoller} L. Jiang, T. Kitagawa, J. Alicea, A. R. Akhmerov, D. Pekker, G. Refael, J. I. Cirac, E. Demler, M. D. Lukin, and P. Zoller, Phys. Rev. Lett. {\bf106}, 220402 (2011).

\bibitem{Tewari} S. Tewari and J. €‰D. Sau, Phys. Rev. Lett. {\bf109}, 150408 (2012).

\bibitem{Yokoyama} M. Sato, Y. Tanaka, K. Yada, T. Yokoyama, Phys. Rev. B {\bf83} 224511 (2011).

\bibitem{Law1} C. L. M. Wong, J. Liu, K. T. Law, and P. A. Lee, Phys. Rev. B {\bf88} 060504(R) (2013).

\bibitem{Spielman3} B. M. Anderson, G. Juzeliunas, V. M. Galitski, and I. B. Spielman, Phys. Rev. Lett. {\bf108}, 235301 (2012).

\bibitem{Juzeliunas} B. M. Anderson, I. B. Spielman, and G. Juzeliunas, Phys. Rev. Lett. {\bf111} 125301 (2013).

\bibitem{Xiongjun} B.-Z. Wang, Y.-H. Lu, W. Sun, S. Chen, Y. Deng and X.-J. Liu, arXiv: 1706.08961.

\bibitem{Shin} Y. €‰I. Shin, C. €‰H. Schunck, A. Schirotzek, and W. Ketterle, Phys. Rev. Lett. {\bf99}, 090403 (2007).

\bibitem{Pu} L. Jiang, L. €‰O. Baksmaty, H. Hu, Y. Chen, and H. Pu, Phys. Rev. A {\bf83}, 061604 (2011).

\bibitem{Mueller} R. Wei and E. €‰J. Mueller, Phys. Rev. A {\bf86}, 063604 (2012).

\bibitem{Baranov} C. V. Kraus, P. Zoller, and M. A. Baranov, Phys. Rev. Lett. {\bf111}, 203001 (2013).

\bibitem{Buchler} A. Buhler, N. Lang, C. V. Kraus, G. Moller, S. D. Huber, and H. P. Buchler, Nat. Commun. {\bf5}, 4504 (2014).

\end{thebibliography}
\end{document}